# CREEPING FLOW OF MICROPOLAR FLUID PARALLEL TO THE AXIS OF CYLINDRICAL CELLS WITH POROUS LAYER

D. Yu. Khanukaeva*[1], A. N. Filippov[1], P. K. Yadav[2], A. Tiwari[3]

[1]*Gubkin Russian State University of Oil and Gas (National Research University)
Leninsky prospect, 65-1, Moscow, 119991, Russia*

[2]*Motilal Nehru National Institute of Technology Allahabad, Allahabad -211004, India*

[3]*Birla Institute of Technology & Science, Pilani-333031, Rajasthan, India*

*e-mail: khanuk@yandex.ru*

**Abstract:** The present paper considers the flow of micropolar fluid through a membrane modeled as a swarm of solid cylindrical particles with porous layer using the cell model technique. Traditional boundary conditions on hypothetical cell surface were added with an additional condition: the no spin condition / no couple stress condition. Expressions for velocity and microrotation vector components have been obtained analytically. Effect of various parameters such as particle volume fraction, permeability parameter, micropolarity number etc. on hydrodynamic permeability of membrane has been discussed.

## 1. Introduction

Flow through random assemblage of particles have always been a topic of interest to researchers due to its vast applications in various real life problems such as flow through sand beds, petroleum reservoirs, membrane filtration processes etc. For modeling of flow through a porous medium, Darcy [1] or Brinkman [2] formulations are used, depending upon the media. Cell model technique is an approach to analyze such problems by taking one particle in the swarm to be confined in a hypothetical cell and applying appropriate boundary condition on the cell surface to include the effect of neighboring particles on the particle concerned.

In this way, the problem of analyzing flow past each particle is reduced to flow past a single particle. Although researchers started working on this area in early fifth decade of the twentieth century, significant improvement came when Happel [3, 4] and Kuwabara [5] took geometry of cell surface and particle to be the same (cylindrical/spherical) and provided vanishing of shear stress and vanishing of rotation at cell surface respectively. Later Mehta-Morse [6] and Kvashnin [7] came up with different conditions on cell surface to analyze flow past swarm of particles: homogeneity of flow and symmetry of the cells, respectively. It is necessary to mention here that the boundary condition to which we usually refer as the Mehta and Morse one, following to J. Happel and H. Brenner, was proposed firstly by Cunningham in 1910 [8]. Perepelkin, Starov, and Filippov [9] presented a solution to the problem of flow past a porous sphere located in the center of a spherical cell. A general solution was found to the Brinkman problem for a porous particle. Solutions were obtained for the case of complete mixing and non-mixing of the liquids inside and outside the particle. Formulas which were found express the force acting on a particle. However in [9] the explicit expression for the hydrodynamic permeability of dispersion was not derived and was calculated by the chain of algebraic formulas. This gap was compensated in [10] where the flow around completely porous particle was studied within the framework of the cell model, and the algebraic expression for the permeability of a set of porous particles (membrane) was derived in the general case of different liquid viscosities inside and outside the porous layer. Moreover, various limiting cases of the flow of viscous liquid around the set of porous particles were studied in [10]. Later in [11], Vasin, Starov and Filippov solved the problem of the motion of partially porous particle consisting of a rigid core, covered with a porous non-deformable hydrodynamically uniform layer, in the unbounded incompressible liquid. A finite algebraic formula was obtained for determining the force that acts on a particle.

Hydrodynamic permeability of a membrane regarded as a system of impermeable particles covered with porous layer was calculated afterwards in [12]. These authors reviewed their investigations [13] of the hydrodynamic permeability of porous membranes built up from completely porous particles or particles with a porous shell using the Cunningham/Mehta-Morse cell condition. Vasin, Starov and Filippov [14] presented a review of a cell method applied for investigations of hydrodynamic permeability of porous/dispersed media and membranes. The hydrodynamic permeability of a porous layer/membrane built up by solid particles with a porous shell and non-porous impermeable interior was calculated. Four known boundary conditions on the outer cell boundary were considered and compared: Happel's, Kuwabara's, Kvashnin's and Cunningham's (usually referred to as Mehta-Morse's condition) model. Vasin and Filippov calculated the hydrodynamic permeability of a membrane simulated by a set of identical impermeable cylinders covered with a porous layer by the Happel–Brenner cell method [15]. Both transverse and longitudinal flows of filtering liquid with respect to the cylindrical fibers that compose the membrane were studied. Boundary conditions on the cell surface that correspond to the Happel, Kuwabara, Kvashnin, and Cunningham models were considered and theoretical values of the Kozeny constant were evaluated. Hydrodynamic permeability of membrane composed of a set of porous spherical particles with rigid impenetrable cores was calculated in [16] on the base of the cell method. All known and abovementioned boundary conditions on the cell surface were considered. Theoretical and empirical results were compared and a good agreement was achieved between them. Different limiting cases for which the derived formulas lead to results known from published literature were also studied. Deo, Yadav and Tiwari [17] discussed flow past a swarm of particles of cylindrical geometry with Happel's formulation and also discussed the effect of various parameters such as particle volume fraction on flow pattern. Deo, Filippov et al. [18] and Yadav, Tiwari et al. [19] compared all four cell models to analyze flow past a swarm of cylindrical/spherical particles and variation of membrane permeability with other parameters by taking stress jump condition at fluid-porous interface which was suggested by Ochoa-Tapia and Whitaker [20, 21]. Vasin et al. [22], Tiwari et al. [23] used Brinkman and Darcy formulations respectively to analyze flow past swarm of solid cylindrical particles with porous layers having varying specific permeability and discussed the effect of various parameters on membrane permeability. Filippov and Koroleva [24] provided a study of a Stokes-Brinkman system with varying viscosity that describes the fluid flow along an ensemble of partially porous cylindrical particles using the cell approach. The existence and uniqueness of the solutions were proved as well as some uniform estimates were derived. Recently Filippov and Ryzhikh [25] employed the cell method to calculate the hydrodynamic permeability of a porous medium (membrane) composed of a set of partially porous spherical particles with solid impermeable cores. This representation is used to describe the globular structure of membranes containing soluble grains. The apparent viscosity of a liquid is suggested to increase as a power function of the depth of the porous shell from the viscosity of the pure liquid at the porous medium–liquid shell interface to some larger value at the boundary with the impermeable core. All known boundary conditions used for the cell surface have been considered.

In all the literature mentioned above, the investigations were done only for flow of Newtonian fluid. However, in many real life problems based on cell model technique, the fluid may be non-Newtonian in nature such as blood flow through arterial wall where animal blood shows characteristic of micropolar fluid. Apart from this, liquid crystals made up of dumbbell molecules, polymeric fluids with additives also behave like micropolar fluids. Eringen [26, 27] used conservation of micro inertia moments and balancing of first stress moments to develop theory of microfluids, which possess local inertia. This theory helped researchers to use inertial spin, body and stress moments, micro stress averages in deriving equations to get complete flow information in specific cases such as polymeric fluids with additives etc. Although micro-fluids take into consideration the micro-rotation and spin inertia but the theory become very complicated for a general class of micro fluids due to the complexities involved in formulations. A micropolar fluid is a subclass of microfluids, which supports couple stress, and body couples only, which makes its formulations relatively simple for researchers. In micropolar fluids, rotation of a fluid point in volume element about centroid is taken into account other than its rigid motion by means of microrotation vector. Ariman et al. [28, 29] gave a detailed review on development of theory and applications of micropolar fluids to real and ideal flow problems. Dey and Mazumdar [30] used micropolar fluids to model oscillatory blood flow through a stenotic artery. Lukaszewicz [31] discussed the mathematical theory of micropolar fluids for stationary and non-stationary problems emphasizing on existence and uniqueness of solutions for above flows with heat convection or diffusion and also discussed its applicability in lubrication theory (journal bearings etc.). Citing similarity in the behavior of various bio-fluids with micropolar fluids such as cervical mucus that is a suspension of macromolecules in water like liquids, Srinivasacharya et al. [32] and Muthu et al. [33] applied micropolar fluid model to analyze peristaltic motion in cylindrical tubes. The list of applications of micropolar theory is not restricted by the cited works. More references can be found in a recent review of Khanukaeva and Filippov [34].

Many authors considered the problem of the micropolar flow over separate bodies of different shapes. But as far as flow of micropolar fluid past a swarm of objects is concerned, only few literatures are available. In the work of Saad [35] cell model is applied to a swarm of fluid spheres. The cases of micropolar core surrounded by viscous shell and vice versa were considered. The development of the cell model to the viscous spheroid in micropolar spheroidal shell is given in the work [36]. Detailed study of cell model applied to the micropolar flow along and perpendicular to cylindrical particles is given in the work of Sherief et al. [37]. The core of the cell was considered to be solid, slip conditions for linear and angular velocities were set on its surface.

Up to the knowledge of the authors, the cell model has not been considered for flow of micropolar liquid past a combined solid-porous core of cylindrical or spherical geometry. A general statement of the problem and various types of boundary conditions were discussed in the review work [34]. The present paper is the first part of the set of three works devoted to the application of cell model with combined solid-porous core involving micropolar fluid flow past them. It considers the flow parallel to the symmetry axis of cylindrical cell. The next paper will be devoted to the perpendicular flow over the cylindrical cell and chaotic configuration of the cells. The third paper will study the flow over a spherical cell.

Two different boundary conditions on the surface of the cell have been employed. For both formulations, the expression for hydrodynamic permeability of swarm has been obtained and effect of various parameters on hydrodynamic permeability is discussed.

## 2. Statement of the problem

A cylindrical cell, shown in Fig.1, consists of a solid impermeable core of radius $a$, coaxial porous layer $a < r < b$ and outer region of free micropolar liquid flow $b < r < c$. The squared relative thickness of porous layer in a cell $b^2 / c^2$ can be chosen to be equal to the volume fraction of particles in the dispersed system (membrane), which is connected to its porosity. The flow is directed along the symmetry axis of the cell and driven by the pressure gradient $-\nabla p$, which is equal to $-\partial p/\partial z$ in the cylindrical coordinate system $(r,\theta,z)$.

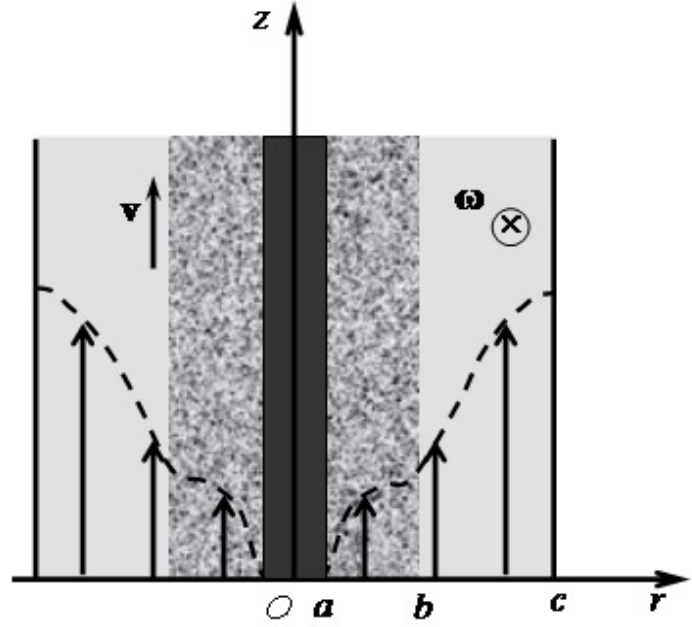

**Fig. 1.** Geometry of the flow.

The micropolar flow is governed by the system of three equations: the continuity equation, the momentum equation and the moment of momentum equation [26, 27]

$$\nabla \cdot \mathbf{v} = 0,$$
$$\rho \dot{\mathbf{v}} = \rho \mathbf{F} - \nabla p + (\mu + \kappa)\Delta \mathbf{v} + 2\kappa \nabla \times \boldsymbol{\omega}, \qquad (1)$$
$$\rho \hat{J}\dot{\boldsymbol{\omega}} = \rho \mathbf{L} + (\alpha + \delta - \varsigma)\nabla \nabla \cdot \boldsymbol{\omega} + (\delta + \varsigma)\Delta \boldsymbol{\omega} + 2\kappa \nabla \times \mathbf{v} - 4\kappa \boldsymbol{\omega},$$

where $\mathbf{v}, \boldsymbol{\omega}$ are linear and angular velocity vectors correspondingly, $\rho$ is the liquid density, $\hat{J}$ the moment of inertia tensor, $\mathbf{F}, \mathbf{L}$ are the densities of external forces and couples, $\mu, \kappa, \alpha, \delta, \varsigma$ are the viscosity coefficients of the micropolar medium. It's worth mentioning that system (1) slightly differs in the notation of coefficients from the original Eringen's formulation, which has been used in the majority of papers. Namely, coefficients $\mu$ and $\kappa$ are introduced in such a way that at $\kappa = 0$ not only the first and the second equations of system (1) are reduced to the Navier-Stokes equations of classical hydrodynamics, but also the viscosity coefficient $\mu$ is equal to a dynamic viscosity of the Newtonian liquid. Detailed consideration of the governing equations formulation and viscosity coefficients definition is given in [34].

For stationary Stokesian flow in absence of external forces and couples velocity and spin vectors have coordinates $\mathbf{v}\{0; 0; u(r)\}$, $\boldsymbol{\omega}\{0; \omega(r); 0\}$. The continuity equation is satisfied automatically. The coordinate form of the rest of the governing equations is ($b < r < c$)

$$(\mu + \kappa)\left(u_2'' + \frac{u_2'}{r}\right) + 2\kappa\left(\omega_2' + \frac{\omega_2}{r}\right) = \frac{\partial p}{\partial z},$$

$$(\delta + \varsigma)\left(\omega_2'' + \frac{\omega_2'}{r} - \frac{\omega_2}{r^2}\right) - 2\kappa u_2' - 4\kappa\omega_2 = 0, \qquad (2)$$

where subscript 2 is being ascribed to all the variables in the free stream layer. Subscript 1 will be used below to represent variables in porous layer. Throughout the manuscript the streak represents derivative with respect to $r$.

Brinkman-type equations will be used for the modeling of the flow in porous region. According to the original Brinkman's idea developed for the stationary filtration flows of non-polar liquids [2], the applied pressure gradient is balanced by the viscous terms and the resistance generated by the porous matrix. The former enters the equation with the coefficient called effective viscosity $\mu'$, which is not necessarily equal to the dynamic viscosity $\mu$. The porous matrix was approximated as a swarm of spherical particles with the drag described by the Stokes formula, which entered the equation with the coefficient $k$ called specific permeability and in fact represents the known Darcy term. Later, in the paper of Ochoa-Tapia and Whitaker [21], the Brinkman equation was rigorously derived by the averaging method over the representative volume. It was shown also that the effective viscosity is equal to $\mu/\varepsilon$, where $\varepsilon$ is the local porosity of the medium (in our case – porous layer surrounding solid core). Thus, the Brinkman equation for Newtonian liquid has the form

$$\nabla p = \frac{\mu}{\varepsilon}\Delta \mathbf{v} - \frac{\mu}{k}\mathbf{v}.$$

Adopting the original Brinkman's idea for the micropolar flow in porous medium, one should write the momentum equation with viscous terms corrected by the coefficient $\varepsilon^{-1}$ and add Darcy-like resistance term calculated for the micropolar flow. Many authors considered the problem of the micropolar flow over the sphere of radius $R$ with various boundary conditions. Rao and Rao [28] took the initiative in this direction and computed the drag force with no-slip and no-spin conditions [38]. It has the most concise expression, and in the notations of the present paper takes the form $F_0 = 6\pi U R \mu \left(1 + \dfrac{\dfrac{\kappa}{\mu}}{1 + 2R\sqrt{\dfrac{\kappa(\mu+\kappa)}{\mu(\delta+\varsigma)}}}\right)$. For the sphere of infinitesimal radius this expression can be reduced to $F = 6\pi U R(\mu + \kappa)$. So, the Darcy-like term for micropolar fluid differs from classical Darcy term by the coefficient $\mu + \kappa$. The moment of momentum equation should remain unchanged, since couple on the sphere is zero. Rigorous mathematical derivation of the equations governing the filtration of micropolar flow is fulfilled in the paper of Kamel et al. [39]. The standard averaging technique over the representative volume is used and the no-slip, no-spin conditions are applied on solid boundaries. The resultant system of equations for the averaged values of the slow stationary flow can be written in the form

$$\nabla \cdot \mathbf{v} = 0,$$

$$\nabla p = \frac{\mu + \kappa}{\varepsilon} \Delta \mathbf{v} + 2\frac{\kappa}{\varepsilon} \nabla \times \boldsymbol{\omega} - \frac{\mu + \kappa}{k} \mathbf{v},$$

$$0 = (\alpha + \delta - \varsigma)\nabla <\nabla \cdot \boldsymbol{\omega}> + (\delta + \varsigma)\Delta \boldsymbol{\omega} + 2\kappa \nabla \times \mathbf{v} - 4\kappa \boldsymbol{\omega},$$

where the average value of the spin divergence over the representative volume $V$ is $<\nabla \cdot \boldsymbol{\omega}> = \frac{1}{V}\int_V \nabla \cdot \boldsymbol{\omega}\, dV$. For the cylindrical geometry and some other practical cases the condition $\nabla \cdot \boldsymbol{\omega} = 0$ is fulfilled, and this term vanishes. Thus, the derived equations are totally coinciding with those which would be obtained by simple mechanical consideration offered by Brinkman.

So, the coordinate form of governing equations for porous region is $(a < r < b)$

$$\frac{\mu + \kappa}{\varepsilon}\left(u_1'' + \frac{u_1'}{r}\right) + 2\frac{\kappa}{\varepsilon}\left(\omega_1' + \frac{\omega_1}{r}\right) - \frac{\mu + \kappa}{k}u_1 = \frac{\partial p}{\partial z},$$

$$(\delta + \varsigma)\left(\omega_1'' + \frac{\omega_1'}{r} - \frac{\omega_1}{r^2}\right) - 2\kappa u_1' - 4\kappa \omega_1 = 0.$$
(3)

General solution for systems (2) and (3) should contain 8 constants, so we need 8 boundary conditions.

On the solid surface it is necessary to set up no-slip and no-spin conditions

$$u_1(a) = 0, \quad \omega_1(a) = 0,$$
(4)

as the filtration equations were derived under this assumption. All kinds of slips are actively used by modern models of micropolar flows, but in order to apply them here one should revise the equations of system (3).

On the fluid-porous surface the most natural conditions seem to be the continuity of velocity and micro rotation vectors respectively

$$u_1(b-0) = u_2(b+0), \quad \omega_1(b-0) = \omega_2(b+0) \tag{5}$$

and continuity of the stress and couple stress tensor components, tangential to the boundary surface. The stress and couple stress tensors [40] using the notations of this paper are correspondingly $\hat{T} = -p\hat{G} + 2\mu\hat{\gamma}^{(S)} + 2\kappa\hat{\gamma}^{(A)}$ and $\hat{M} = \alpha(\mathrm{tr}\hat{\chi})\hat{G} + 2\delta\hat{\chi}^{(S)} + 2\varsigma\hat{\chi}^{(A)}$, where $\hat{G}$ is the metric tensor, superscripts (S) and (A) denote correspondingly symmetric and skew-symmetric parts of tensors, $\hat{\gamma} = (\nabla \mathbf{v})^T - \hat{\varepsilon} \cdot \boldsymbol{\omega}$, $\hat{\varepsilon}$ is the Levi-Civita tensor, $\hat{\chi} = (\nabla \boldsymbol{\omega})^T$, superscript $T$ stands for transposed tensors, the gradients of linear and angular velocity tensors $\nabla \mathbf{v}$ and $\nabla \boldsymbol{\omega}$ defined in Cartesian coordinates have components $(\nabla \mathbf{v})_{ij} = \partial v_i / \partial x_j$, $(\nabla \boldsymbol{\omega})_{ij} = \partial \omega_i / \partial x_j$. In the chosen cylindrical coordinate system the components of the stress and couple stress tensors are

$$\hat{T} = \begin{pmatrix} -p & 0 & (\mu+\kappa)u'(r) + 2\kappa\omega(r) \\ 0 & -pr^2 & 0 \\ (\mu-\kappa)u'(r) - 2\kappa\omega(r) & 0 & -p \end{pmatrix},$$

$$\hat{M} = \begin{pmatrix} 0 & (\delta+\varsigma)\omega'(r) - (\delta-\varsigma)\dfrac{\omega(r)}{r} & 0 \\ (\delta-\varsigma)\omega'(r) - (\delta+\varsigma)\dfrac{\omega(r)}{r} & 0 & 0 \\ 0 & 0 & 0 \end{pmatrix}.$$

The required components are $T_{13}$, $M_{12}$ correspondingly.

In porous region all the viscous terms should be divided by the local porosity as it was mentioned above. Using this notation, we can write the boundary condition for stresses and couple stresses as

$$\frac{\mu+\kappa}{\varepsilon}u'_1(b-0) + 2\frac{\kappa}{\varepsilon}\omega_1(b-0) = (\mu+\kappa)u'_2(b+0) + 2\kappa\omega_2(b+0),$$
$$\frac{\delta+\varsigma}{\varepsilon}\omega'_1(b-0) - \frac{\delta-\varsigma}{\varepsilon}\frac{\omega_1(b-0)}{b-0} = (\delta+\varsigma)\omega'_2(b+0) - (\delta-\varsigma)\frac{\omega_2(b+0)}{b+0}. \tag{6}$$

In order to close the boundary value problem two conditions are required at $r=c$. In classical cell models for non-polar liquids there are four known types of conditions for the outer boundary of the cell. They are Happel's [3, 4], Kuwabara's [5], Kvashnin's [7], and Mehta-Morse [6]/Cunningham [8] conditions. For micropolar liquid all of them can be considered provided that each is supplemented with one more condition. Here we use Happel's no-stress condition

$$(\mu+\kappa)u'_2(c-0) + 2\kappa\omega_2(c-0) = 0 \tag{7}$$

along with no-couple stress condition

$$(\delta+\varsigma)\omega'_2(c-0) - (\delta-\varsigma)\frac{\omega_2(c-0)}{c-0} = 0, \tag{8}$$

as suggested by Sherief et al. [37], where a cell did not contain porous layer. As an alternative to no-couple stress condition, we use the vanishing spin condition

$$\omega_2(c-0) = 0, \tag{9}$$

as considered by Saad [35]. Symmetry condition for micro-rotation or some relation between spin and flow vorticity or something else can be set on the surface of the cell. There is no obvious evidence of the global preference of any of four known conditions for non-polar flows. Even less data available on the conditions for polar liquids. All combinations of four mentioned conditions with new ones are possible. This is a theme for a future separate investigation. In this paper we compare solutions of systems (2) and (3) with boundary conditions (4-7) and (8) or (9). The governing equations of both boundary value problems (BVPs) have been solved analytically.

## 3. General solution of the problem

Let us introduce non-dimensional variables and parameters:

$$\tilde{r} = \frac{r}{b},\ \tilde{z} = \frac{z}{b},\ \ell = \frac{a}{b},\ m = \frac{c}{b},\ \tilde{u} = \frac{u\rho b}{\mu},\ \tilde{\omega} = \frac{\omega\rho b^2}{\mu},\ \frac{\partial \tilde{p}}{\partial \tilde{z}} = \frac{\partial p}{\partial z}\frac{\rho b^3}{\mu^2}. \qquad (10)$$

We adopt the usual notations in micropolar theory for some important parameters

$$N^2 = \frac{\kappa}{\mu + \kappa};\quad \tau^2 = \frac{1}{4}\frac{\delta + \varsigma}{\mu};\quad L^2 = \frac{\tau^2}{b^2}. \qquad (11)$$

The so-called micropolarity number $N$ is connected to the dynamic micro-rotation viscosity $\kappa$. It characterizes the measure of effectiveness for the angular momentum transmission between the medium particles. Classical limit corresponds to $N = 0$, while maximal value $N = 1$ corresponds to a strong micropolarity. Characteristic length $\tau$ plays a role of microscale for the medium and sometimes is interpreted as "particles" size. The bigger the value of $\tau$, the stronger the polar effects are exhibited. One should remember that this parameter is not a real size of any physical object and formally can have any value, allowed by the constraints on the viscosity coefficients [40]:

$$\mu \geq 0;\quad \kappa \geq 0;$$
$$\delta \geq 0,\ \delta + \varsigma \geq 0,\ 3\alpha + 2\delta \geq 0,\ -(\delta + \varsigma) \leq \delta - \varsigma \leq \delta + \varsigma.$$

Thus, non-dimensional parameters $N$ and $L$, defined by relations (11), have their values in the intervals $N \in [0;\ 1);\ L \in [0;\ \infty)$. They will be used during parametric studies of the solutions.

In notations (11) the non-dimensional form of system (2) reduces to $(1 < r < m)$

$$\frac{1}{1-N^2}\left(u_2'' + \frac{u_2'}{r}\right) + 2\frac{N^2}{1-N^2}\left(\omega_2' + \frac{\omega_2}{r}\right) = \frac{\partial p}{\partial z},$$

$$L^2\frac{1-N^2}{N^2}\left(\omega_2'' + \frac{\omega_2'}{r} - \frac{\omega_2}{r^2}\right) - \frac{1}{2}u_2' - \omega_2 = 0. \quad (12)$$

Tilde is omitted here and further. System (3) for porous region takes the following non-dimensional form ($\ell < r < 1$)

$$\frac{1/\varepsilon}{1-N^2}\left(u_1'' + \frac{u_1'}{r}\right) + 2\frac{N^2/\varepsilon}{1-N^2}\left(\omega_1' + \frac{\omega_1}{r}\right) - \frac{\sigma^2}{1-N^2}u_1 = \frac{\partial p}{\partial z},$$

$$L^2\frac{1-N^2}{N^2}\left(\omega_1'' + \frac{\omega_1'}{r} - \frac{\omega_1}{r^2}\right) - \frac{1}{2}u_1' - \omega_1 = 0, \quad (13)$$

where $\sigma = b/\sqrt{k}$ characterizes the ratio of macro and micro scales of porous medium, as $\sqrt{k}$ has the dimension of length and is often called Brinkman's radius.

General solution of system (12) contains four arbitrary constants $C_1, C_2, C_3, C_4$ and includes modified Bessel functions $I_0(rN/L)$, $I_1(rN/L)$ and Mcdonald functions $K_0(rN/L)$, $K_1(rN/L)$ of the zero and the first order correspondingly:

$$u_2(r) = C_1 + \frac{1}{4}\frac{\partial p}{\partial z}r^2 + C_2\ln r - C_3 I_0(rN/L) - C_4 K_0(rN/L), \quad (14)$$

$$\omega_2(r) = -\frac{1}{4}\frac{\partial p}{\partial z}r - \frac{C_2}{2r} + \frac{1}{2NL}\left(C_3 I_1(rN/L) - C_4 K_1(rN/L)\right). \quad (15)$$

General solution of system (13) also includes modified Bessel and Mcdonald functions and arbitrary constants $C_5, C_6, C_7, C_8$:

$$u_1(r) = -\frac{1-N^2}{\sigma^2}\frac{\partial p}{\partial z} + C_5 I_0(\alpha_1 r) + C_6 K_0(\alpha_1 r) + C_7 I_0(\alpha_2 r) + C_8 K_0(\alpha_2 r), \quad (16)$$

$$\omega_1(r) = \frac{1}{2N^4}\Big(\alpha_1\big[-N^4 + L^2(1-N^2)(\varepsilon\sigma^2 - \alpha_1^2)\big]\big(C_5 I_1(\alpha_1 r) - C_6 K_1(\alpha_1 r)\big) + $$
$$+ \alpha_2\big[-N^4 + L^2(1-N^2)(\varepsilon\sigma^2 - \alpha_2^2)\big]\big(C_7 I_1(\alpha_2 r) - C_8 K_1(\alpha_2 r)\big)\Big), \quad (17)$$

where $\alpha_1$ and $\alpha_2$ are defined by the system $\begin{cases} \alpha_1^2 + \alpha_2^2 = \sigma^2\varepsilon + N^2/L^2, \\ \alpha_1^2 \alpha_2^2 = \dfrac{\sigma^2 \varepsilon}{1-N^2}\dfrac{N^2}{L^2}. \end{cases}$

## 4. Solution of the boundary value problems

Boundary conditions (4)-(8) in the non-dimensional form are as follows:

$$u_1(\ell) = 0, \quad \omega_1(\ell) = 0, \tag{18}$$

$$u_1(1-0) = u_2(1+0), \quad \omega_1(1-0) = \omega_2(1+0), \tag{19}$$

$$\frac{1}{\varepsilon N^2}u_1'(1-0) + \frac{2}{\varepsilon}\omega_1(1-0) = \frac{1}{N^2}u_2'(1+0) + 2\omega_2(1+0), \tag{20}$$

$$\frac{1}{\varepsilon}\omega_1'(1-0) - \frac{\phi}{\varepsilon}\omega_1(1-0) = \omega_2'(1+0) - \phi\omega_2(1+0), \tag{21}$$

$$\frac{1}{N^2}u_2'(m-0) + 2\omega_2(m-0) = 0, \tag{22}$$

$$\omega_2'(m-0) - \phi\frac{\omega_2(m-0)}{c-0} = 0. \tag{23}$$

Non-dimensional form of condition (9) is

$$\omega_2(m-0) = 0. \tag{24}$$

One can notice, that conditions (6) and (8) contain viscosities $\delta$ and $\varsigma$ in the form non reducible to two introduced combinations $N$ and $L$. This is the consequence of non-symmetry of the couple stress tensor. The only simplification which can be done is the reduction to one additional parameter $\phi = (\delta - \varsigma)/(\delta + \varsigma)$, which according to the constraints on the viscosity coefficients can vary in the interval $\phi \in [-1;\ 1]$. The presence of viscosities combination in boundary conditions indicates that properties of the micropolar medium are maximally exhibited at the boundaries. In addition, the case of $\delta = \varsigma$ ($\phi = 0$) is of great importance, as it simplifies the conditions (21) and (23) making these boundary conditions and consequently the solution of the problem independent of $\delta$ and $\varsigma$ explicitly and again dependent only on two parameters $N$ and $L$. So, consideration and discussion of various boundary conditions for micropolar flows are of greatest importance.

The characteristic profiles of solutions (14)-(17) with the conditions (18)-(24) are shown in Fig.2 for the following values of parameters: $\partial p/\partial z = -1$, $N = 0.5$, $L = 0.2$, $\ell = 0.5$, $m = 1.5$, $\varepsilon = 0.75$, $\sigma = 3$, $\phi = 0.5$. Each of the chosen magnitudes represents value of the parameter almost in the middle of the allowed limits. The values assumed for parameters $N$, $L$ and $\phi$ correspond to well-developed micropolarity, and the chosen values of $\varepsilon$ and $\sigma$ characterize a porous medium, exhibiting its properties significantly. The fractions of filtration and micropolar flows are assumed to have equal widths and pressure gradient equal to minus unity. The choice of non-dimensionalization for the linear velocity allow us to treat its non-dimensional value as the Reynolds number (see formulas (10)). So, the scale of ordinate in Fig.2a confirms the validity of the Stokes approximation. In all the graphical analysis involving effect of a parameter on hydrodynamic permeability, the remaining parameters assume only above listed values.

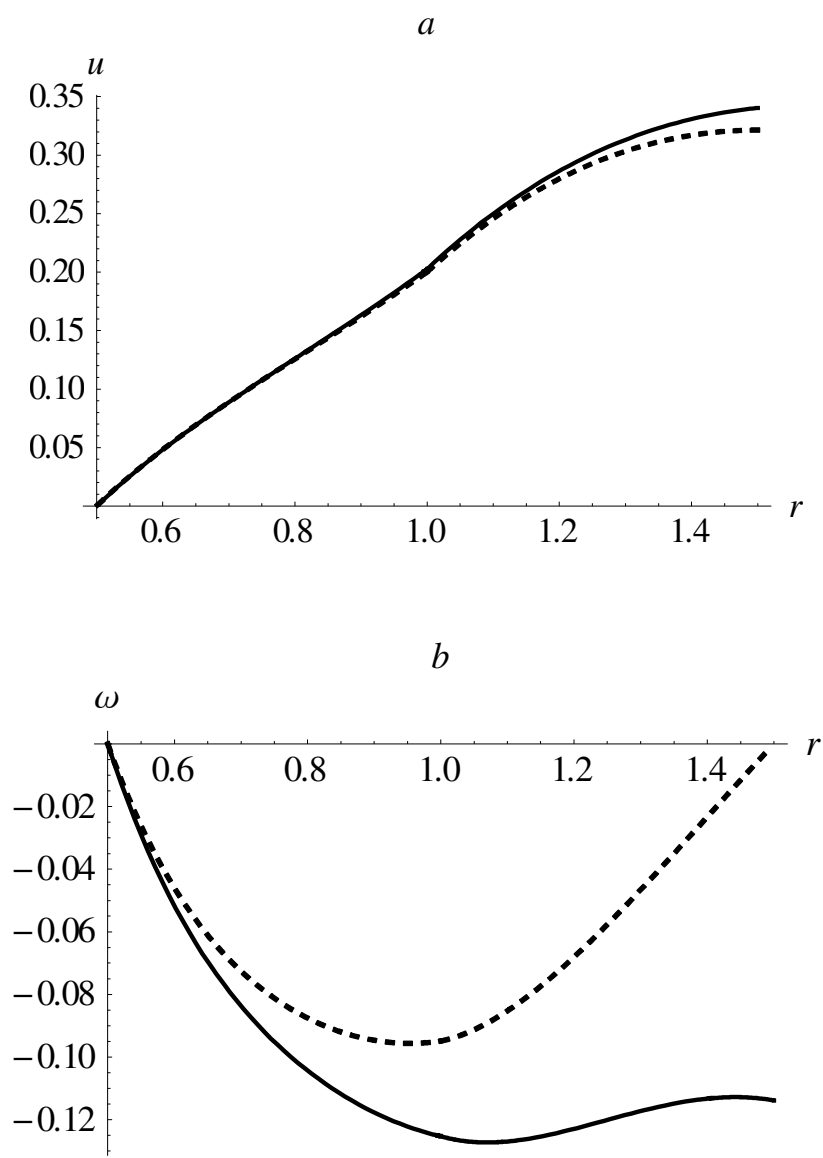

**Fig.2.** Variation of linear velocity (a) and angular velocity (b) under no couple stress condition (solid lines) and no spin condition (dashed lines) on the outer surface of the cell.

Fig.2-Fig.8 contain solid and dashed curves. Solid curve corresponds to the solution of the problem with no couple stress condition (23), dashed line represents the results for the no spin condition (24). As the considered BVPs statements have the only difference on the cell surface and this difference involves spin vector, the biggest discrepancy in curves is observed for the angular velocities at $r \to m$. Dashed curve in Fig.2b tends to zero with $r \to m$ in accordance with condition (24), while solid curve in Fig.2b demonstrates behavior analogous to the processes with free boundary for $r \to m$, which is characterized by no couple stress condition (23). Negative values of angular velocity in Fig.2b point out that the direction of the spin vector is opposite to that shown in Fig.1. Linear velocity profiles, shown in Fig.2a, increase from solid core towards cell surface under both boundary conditions (23) and (24). They demonstrate very close to each other and almost linear growth in the porous region and deformed parabolic growth in the liquid region. The difference of boundary conditions on the outer cell surface is responsible for the discrepancy in curves at $r \to m$. One can notice that lower values of angular velocity modulus, corresponding to condition (24), result in lower values of linear velocity.

## 5. Results and discussion

The coefficient of hydrodynamic permeability $L_{11}$ is chosen as the characteristics of the dispersed medium (membrane) which will be analyzed further.

The dimensionless flow rate of fluid flowing through the cylinder is

$$Q = 2\pi \left( \int_{\ell}^{1} u_1(r) r dr + \int_{1}^{m} u_2(r) r dr \right).$$

Also, the filtration velocity is $V_f = Q/(\pi m^2)$. Applying the Darcy law for the whole membrane as to the porous medium one obtains the non-dimensional form of its hydrodynamic permeability as $L_{11} = -\dfrac{V_f}{\partial p / \partial z}$. By definition it does not depend on pressure gradient. Effect of all the rest parameters of the problem on hydrodynamic permeability are shown below.

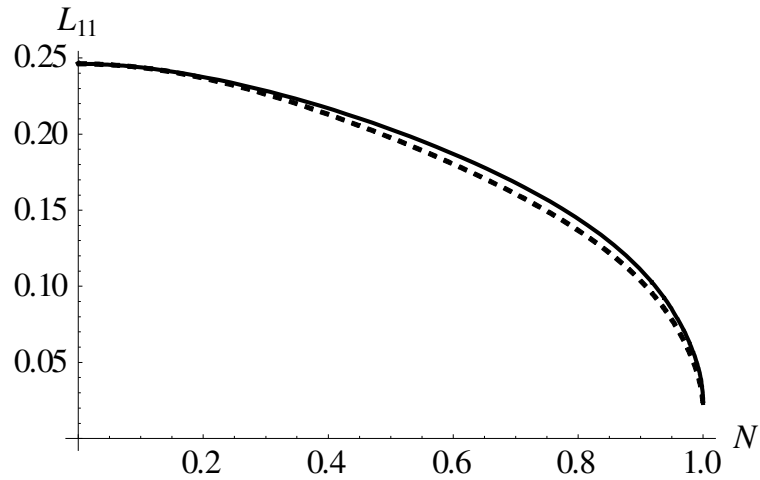

**Fig.3.** Variation of hydrodynamic permeability with coupling parameter *N* under no couple stress condition (solid line) and no spin condition (dashed line) on the outer surface of the cell.

The most interesting is the effect of micropolar properties of the liquid on the flow. It is defined by parameters $N$, $L$, $\phi$. Fig.3 shows the dependency of $L_{11}$ on $N$ for both problem formulations, i.e. using solutions (16)-(17) with conditions (18)-(22) and (23) or (24). It is observed a monotonous and significant decay of hydrodynamic permeability with the growth of coupling parameter in both cases. The increase of coupling is equivalent to the increase of the coefficient of microrotation viscosity, i.e. micro level effects; this leads to the slowing down the flow and, hence, the dropping down of the hydrodynamic permeability of the membrane. The location of dashed line below the solid line in Fig.3 is due to the same reason - the values of flow velocities are less for the BVP with no spin condition in comparison with the velocities for the BVP with no couple stress condition as it can be seen in Fig.2a.

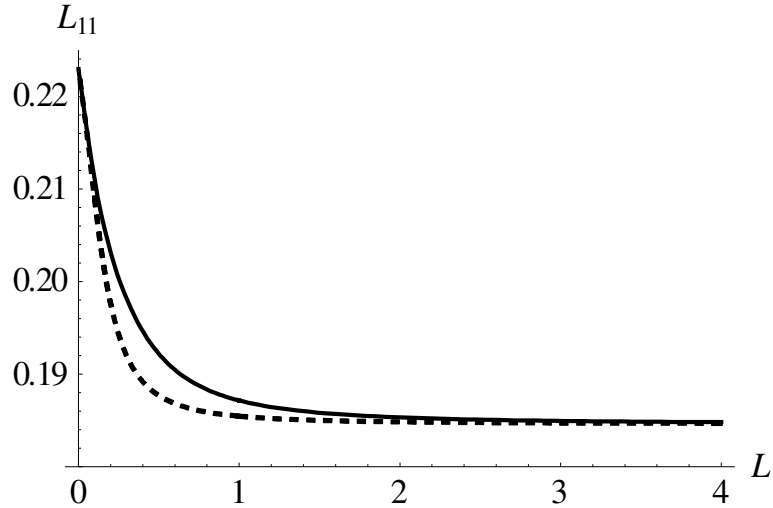

**Fig.4.** Variation of hydrodynamic permeability with scale parameter $L$ under no couple stress condition (solid line) and no spin condition (dashed line) on the outer surface of the cell.

Fig.4 represents the dependency of $L_{11}$ on scale parameter $L$ for both considered BVPs. The hydrodynamic permeability of membrane decreases with parameter $L$ under both boundary conditions and demonstrates asymptotic behaviour. The value of asymptote depends on parameter $N$ and the rate of decrease i.e. the position of initial point of asymptotic behavior is also connected with $N$: for less $N$ asymptotic value of $L_{11}$ is higher and is reached earlier. This means the higher the coupling effect the more important role the particle size plays. And the increase of micro-scale corresponding to a strong polar effect leads towards significant decay of the membrane permeability.

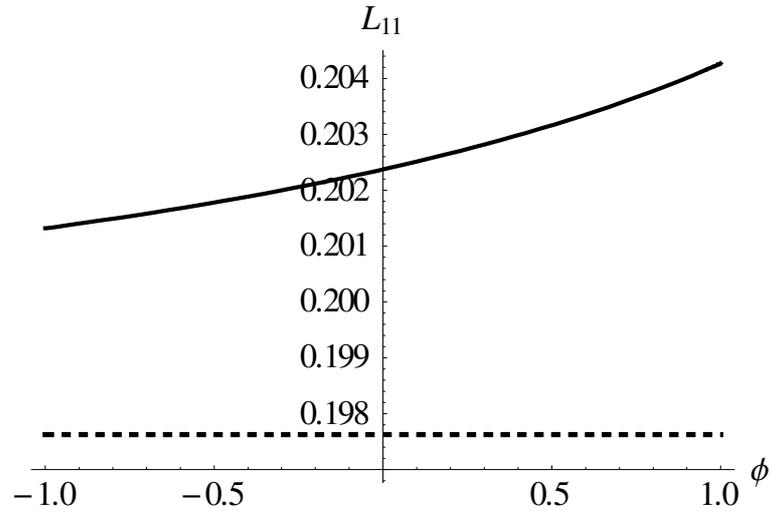

**Fig.5.** Variation of hydrodynamic permeability with parameter $\phi$ under no couple stress condition (solid line) and no spin condition (dashed line) on the outer surface of the cell.

Fig.5 shows the dependency of hydrodynamic permeability of the medium on $\phi$ for the BVP with no couple stress condition on the outer boundary of the cell. Parameter $\phi$ does not take part in the BVP formulation with no spin condition on the outer surface of the cell, but constant value of $L_{11}$ for this BVP is also shown for the sake of comparison. Like the other parametric studies, here also a relatively lower value of $L_{11}$ for no couple stress condition is observed due to the reason stated in the beginning of this section. Monotonous increase of hydrodynamic permeability with $\phi$ signifies different roles of angular viscosities $\delta$ and $\varsigma$ and consequently, the symmetrical and skew symmetrical parts of the curvature-twist rate tensor in couple stresses. Higher relative weight of symmetrical part leads to the better hydrodynamic permeability. This fact additionally confirms more general conclusion about micropolar liquids, that the dominant exhibition of non-symmetry in the medium leads to reduced flow rate characteristics.

Nevertheless, the relative variation of $L_{11}$ does not exceed 2% for the whole range of $\phi$ variation. Besides, the relative difference of the mean value of $L_{11}$ for the BVP with no couple stress condition from its constant value for the BVP with no spin condition is no more than 3%, all the rest parameters being kept constant as listed above. For the same conditions relative variation of $L_{11}$ with parameter $L$ is estimated as 20%, while its variation with parameter $N$ reaches almost 400%. Moreover, variation of both $L$ and $N$ in their full ranges of values gives more than 40 times variation of $L_{11}$.

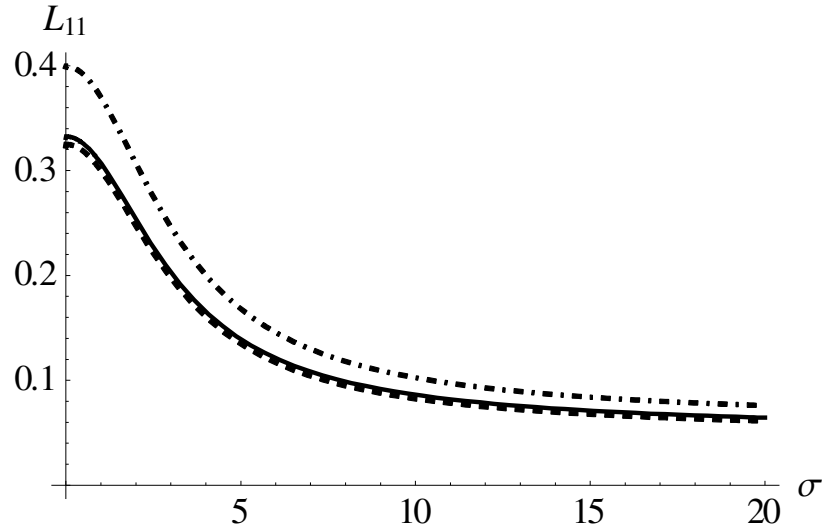

**Fig.6.** Variation of hydrodynamic permeability with parameter σ under no couple stress condition (solid line), no spin condition (dashed line) on the outer surface of the cell and for the Newtonian liquid (dot-dashed line).

The influence of porous layer parameters ε and σ on the hydrodynamic permeability demonstrates a behavior analogous to the case of non-polar liquids investigated in the paper of Vasin [41]. The curve of the hydrodynamic permeability variation with σ for the Newtonian flow in the same cell is shown in Fig.6 with dot-dashed line. It is located obviously higher than analogous curves for polar liquid, because polar flow is usually slower than non-polar one except for the cases of some special types of boundary conditions not discussed here. The scale parameter σ can be associated with the filtration characteristics of porous medium. For normal conditions, it should be sufficiently higher than unity, limit case of $\sigma \to 0$ corresponds to the liquid layer $\ell < r < 1$ with the viscosities, equal to the viscosities in layer $1 < r < m$, divided by $\varepsilon$. Limit case of infinite values of $\sigma$ corresponds to the impermeable layer $\ell < r < 1$, so that the volume of flow reduces to the layer $1 < r < m$. These notes totally correlate with the decrease of hydrodynamic permeability with $\sigma$, shown in Fig.6.

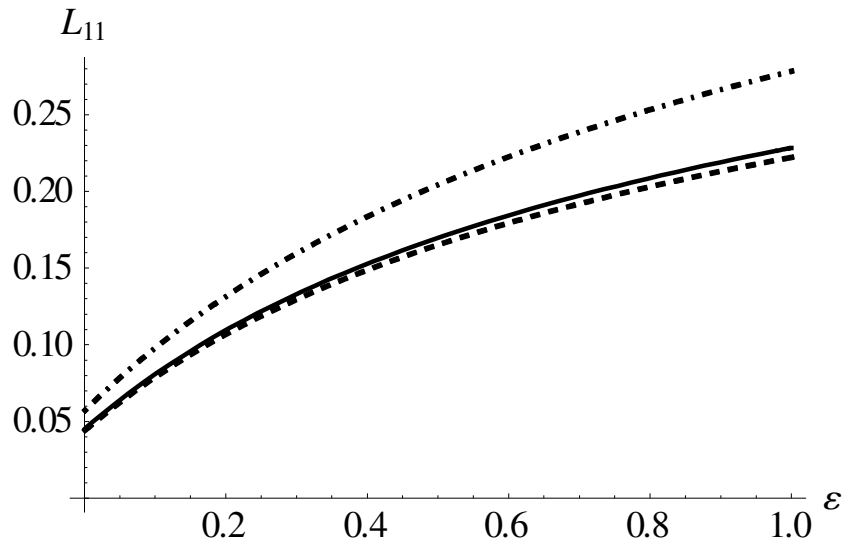

**Fig.7.** Variation of hydrodynamic permeability with porosity ε under no couple stress condition (solid line), no spin condition (dashed line) on the outer surface of the cell and for the Newtonian liquid (dot-dashed line).

The growth of porosity ε of the layer $\ell < r < 1$ obviously results in the increase of the hydrodynamic permeability, shown in Fig.7. Although the curves in Fig.7 are plotted for the whole range of possible values of porosity, one should remember that the validity of Brinkman's approach used in this work is proved only for $\varepsilon > 0.6$. Also the corresponding curve for non-polar liquid is shown in Fig.7 for the sake of comparison.

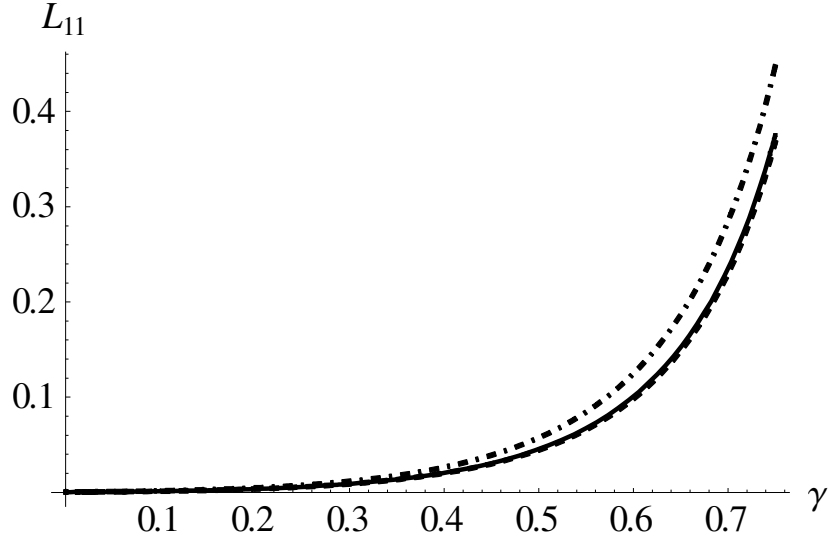

**Fig.8.** Variation of hydrodynamic permeability with porosity γ under no couple stress condition (solid line), no spin condition (dashed line) on the outer surface of the cell and for the Newtonian liquid (dot-dashed line).

Apart from the intrinsic porosity ε of the layer there is usually considered the apparent porosity γ of the membrane as a whole, defined as follows $\gamma = \dfrac{c^2 - b^2}{c^2} = 1 - \dfrac{1}{m^2}$. Strictly speaking the volume of pores in the cell should be calculated taking into account the space occupied by pores of the layer $\ell < r < 1$. In this case the apparent porosity of the membrane $\gamma_\varepsilon$ depends on the intrinsic porosity ε and both geometrical parameters $\ell$ and $m$ and has the form $\gamma_\varepsilon = \dfrac{c^2 - b^2 + \varepsilon(b^2 - a^2)}{c^2} = 1 - \dfrac{1}{m^2} + \varepsilon\dfrac{1-\ell^2}{m^2}$. In case of thin porous layer inside a cell ($\ell \to 1$) porosity $\gamma_\varepsilon$ tends to γ, which depends on the only parameter *m* and allows simple parametric study of the hydrodynamic permeability on γ. The variation of $L_{11}$ with γ is shown in Fig.8 for $\ell = 0.9$ and all the rest parameters assume values as listed above. Fig.8 demonstrates high sensitivity of hydrodynamic permeability to the change of membrane porosity, which can be easily modeled by the geometry of the cell. The same tendency is observed for the Newtonian liquid as it can be seen from Fig.8.

## 6. Conclusion

The present work is a novel concept of applying cell model technique using a non-Newtonian fluid approach to analyze flow along a swarm of partly porous cylindrical particles. The hydrodynamic permeability of the swarm (membrane) was taken as the integral characteristics of the flow. The dependence of the hydrodynamic permeability on the effects of microlevel properties of the liquid, porous properties of the membrane and the boundary conditions on the outer surface of the cell have also been studied.

It has been observed that hydrodynamic permeability is highly dependent on the porosity of the membrane, both apparent and intrinsic, as well as the scale characteristics of the porous layer included in the cell. Micro-level properties of fluid such as coupling parameter and micropolar scale coefficient affect hydrodynamic permeability nearly as significantly as the characteristics of porous medium. The variation of angular viscosities entering one of the considered BVP almost does not affect hydrodynamic permeability.

Also, somewhat unexpected result corresponds to very weak dependence of hydrodynamic permeability on the boundary conditions at the outer surface of the cell. The reason lies in the particular construction of the conditions used. Both of them dealt with the microrotation aspects, while the hydrodynamic permeability chosen as integral characteristics for the analysis includes only linear velocity of the flow. So, microproperties on the boundary may be insufficient for the flow as a whole. Nevertheless, this conclusion should be checked and confirmed for other types of boundary conditions and integral characteristics of the flow. This is the scope of future investigations. This paper is only the first attempt in the analysis of these relations.

Due to the sharp lack of experimental works on the properties of micropolar liquids, especially, flowing through porous media, theoretical studies, such as the presented one, are of highest importance. It gives an insight of facts to be taken into account while processing filtration involving polymeric fluids with additives (dopants) and many other such fluids showing similarity with micropolar fluids.

**Acknowledgements:**

The present work is supported by DST (INT/RUS/ RFBR/ P-212) for Indian side and RFBR (15-58-45142_IND and 17-08-01287) for Russian side.